\documentclass{aa}
\usepackage{txfonts}
\usepackage{graphicx}
\usepackage{natbib}
\bibpunct{(}{)}{;}{a}{}{,}
\usepackage[english]{babel}
\usepackage{aalongtable}
\begin{document}
\title{XMM-Newton spectroscopy \\ of the metal depleted T~Tauri star \object{TWA~5}}
\author{C.~Argiroffi\inst{1} \and A.~Maggio\inst{2} \and G.~Peres\inst{1} \and B.~Stelzer\inst{1, 2} \and R.~Neuh\"auser\inst{3}}
\offprints{C. Argiroffi}
\institute{Dipartimento di Scienze Fisiche ed Astronomiche, Sezione di Astronomia, Universit\`a di Palermo, Piazza del Parlamento 1, 90134 Palermo, Italy, \email{argi@astropa.unipa.it, peres@astropa.unipa.it, stelzer@astropa.unipa.it} \and INAF - Osservatorio Astronomico di Palermo, Piazza del Parlamento 1, 90134 Palermo, Italy, \email{maggio@astropa.unipa.it} \and Astrophysikalisches Institut und Universit\"ats-Sternwarte, Schillerg\"asschen 2-3, D-07745 Jena, Germany, \email{rne@astro.uni-jena.de}}
\date{Received 20 January 2005/ Accepted 28 April 2005}
\authorrunning{C. Argiroffi et al.}
\titlerunning{XMM-Newton spectroscopy of TWA~5}

\abstract{We present results of X-ray spectroscopy for \object{TWA~5}, a member of the young TW~Hydrae association, observed with {\it XMM-Newton}. \object{TWA~5} is a multiple system which shows H$\alpha$ emission, a signature typical of classical T~Tauri stars, but no infrared excess. From the analysis of the RGS and EPIC spectra, we have derived the emission measure distribution vs. temperature of the X-ray emitting plasma, its abundances, and the electron density. The characteristic temperature and density of the plasma suggest a corona similar to that of weak-line T~Tauri stars and active late-type main sequence stars. \object{TWA~5} also shows a low iron abundance ($\sim 0.1$ times the solar photospheric one) and a pattern of increasing abundances for elements with increasing first ionization potential reminiscent of the inverse FIP effect 
observed in highly active stars. The especially high ratio ${\rm Ne/Fe}\sim10$ is similar to that of the classical T~Tauri star \object{TW~Hya}, where the accreting material has been held responsible for the X-ray emission. We discuss the possible role of an accretion process in this scenario. Since all T~Tauri stars in the TW Hydrae association studied so far have very high ${\rm Ne/Fe}$ ratios, we also propose that environmental conditions may cause this effect.
}
\maketitle
\section{Introduction}

T~Tauri stars are young late-type stars with an age of a few Myr, contracting toward the zero age main sequence phase \citep[][and references therein]{FeigelsonMontmerle1999}. They are classified in two groups: classical T~Tauri stars (CTTSs) and weak-line T~Tauri stars (WTTSs). This classification is based on  ${\rm H}\alpha$ emission. CTTSs show strong H$\alpha$ emission (${\rm EW}>10$\,\AA). They are still accreting material from their circumstellar disk, and broad and asymmetric ${\rm H}\alpha$ emission is a direct evidence of this process. In WTTSs H$\alpha$ emission is less strong, indicating that the accretion process has ended and the star is approaching the main-sequence. In most cases CTTSs are also characterized by an infrared excess which marks the presence of a circumstellar disk. The infrared excess is usually considered a prerequisite for accretion, but it does not imply that accretion actually takes place; in fact, some WTTSs also show an infrared excess although much fainter than in CTTSs. Since coeval CTTSs and WTTSs are often observed in the same star forming region, the duration of the accretion phase appears to be different from star to star.

\begin{table*}
\begin{center}
\caption{Characteristics of the sample of T~Tauri stars, sorted by H$\alpha$ emission, for which high resolution X-ray spectra have been analyzed. In case of resolved multiple systems the stellar mass, spectral type, and bolometric luminosity refer to the component responsible for the X-ray emission.}
\label{tab:stars}
\begin{tabular}{lccrcccc}
\hline
\hline
Name               & Mass          & Spectral & \multicolumn{1}{c}{EW(H$\alpha$)$^{\rm a}$} & $L_{\rm X}^{\rm b}$   & $\log (L_{\rm X}^{\rm b}/L_{\rm bol})$ & $N_{\rm e}^{\rm c}$   & References$^{\rm d}$ \\
                   & ($M_{\odot}$) & Type     & \multicolumn{1}{c}{(\AA)}                   & ${\rm (erg\,s^{-1})}$ &                                        & ${\rm (cm^{-3})}$       &            \\
\hline
\object{TW Hya}    & $\sim0.7$     & K7       & -220.0 \hspace{0.01\textwidth}              & $1.3\times10^{30}$    & -2.7                                   & $\sim10^{13}$         & 1, 2, 3, 4 \\
\object{TWA 5}     & $\sim0.5$     & M1.5     &  -13.4 \hspace{0.01\textwidth}              & $6.7\times10^{29}$    & -3.1                                   & $<10^{11}$            & 2, 5, 6    \\
\object{HD 98800}  & $\sim1.1$     & K5       &    0.0 \hspace{0.01\textwidth}              & $4.1\times10^{29}$    & -3.8                                   & $\lesssim10^{11}$     & 2, 7, 8    \\
\object{PZ Tel}    & $\sim1.1$     & K0       &    0.1 \hspace{0.01\textwidth}              & $2.2\times10^{30}$    & -3.2                                   & $<10^{12}$            & 9, 10, 11   \\
\object{HD 283572} & $\sim1.8$     & G5       &    1.1 \hspace{0.01\textwidth}              & $7.8\times10^{30}$    & -3.1                                   & $\cdot\cdot\cdot$     & 12, 13, 14 \\
\hline
\end{tabular}
\end{center}
$^{\rm a}$~Negative values of H$\alpha$ equivalent width mark an emission line. $^{\rm b}$~X-ray luminosity evaluated in the $6-20$\,\AA~band, using the {\it XMM}/MOS or {\it Chandra}/HETGS best fit models presented in the relevant papers. $^{\rm c}$~Densities estimated from the \ion{O}{vii} and \ion{Ne}{ix} triplets. $^{\rm d}$ Data from: (1) \citet{BatalhaBatalha2002}; (2) \citet{Reid2003}; (3) \citet{StelzerSchmitt2004}; (4) \citet{KastnerHuenemoerder2002}; (5) this work; (6) \citet{JensenCohen1998}; (7) \citet{KastnerHuenemoerder2004}; (8) \citet{PratoGhez2001}; (9) \citet{CutispotoPastori2002}; (10) \citet{ThatcherRobinson1993}; (11) \citet{ArgiroffiDrake2004}; (12) \citet{StrassmeierRice1998}; (13) \citet{FernandezMiranda1998}; (14) \citet{ScelsiMaggio2005}.

\end{table*}

One of the signatures of stellar youth is a high X-ray emission level. Many star forming regions have been under investigation in order to infer the properties of X-ray emission from pre-main-sequence (PMS) stars. One of the debated questions is whether and how the X-ray emission of accreting CTTSs and non accreting WTTSs differs. It is conceivable that the occurrence of the accretion process in CTTSs might play a role in determining the different X-ray emission characteristics. In fact, the circumstellar disk is thought to affect the geometry of the stellar magnetosphere \citep{Koenigl1991,BouvierGrankin2003}. Moreover accreting material may provide an alternative heating mechanism for the emitting plasma, although shock heated plasma cannot attain temperatures higher than a few MK. The picture emerging from the analysis of low resolution X-ray spectra of PMS stars is that the X-ray luminosity of CTTSs is lower than that of WTTSs, and the X-ray spectra produced by CTTSs appear harder than WTTS spectra \citep{NeuhaeuserSterzik1995,StelzerNeuhaeuser2000,TsujimotoKoyama2002,FlaccomioMicela2003,StassunArdila2004,OzawaGrosso2005}. The harder X-ray spectra of CTTSs may be explained with the presence of plasma hotter ($T\sim10-100$\,MK) than that of WTTSs \citep[$T\sim10$\,MK, ][]{TsujimotoKoyama2002}. If this is the case, the shock heating mechanism cannot be responsible for the X-ray emission in CTTSs. However, it is also possible that circumstellar material absorbs the softest part of the X-ray radiation, simulating therefore a higher temperature in CTTSs \citep{StassunArdila2004}.

\begin{table*}
\begin{center}
\caption{Log of the {\it XMM} observation of \object{TWA~5} (Rev. 565, ObsId 0112880101).}
\label{tab:log}
\begin{tabular}{lccccc}
\hline\hline
Instrument & Science      & Filter  & Start               & Exposure & Count Rate      \\
           & Mode         &         & (UT)                & (ks)     & (${\rm cts\,s^{-1}}$) \\
\hline
PN         & Full Frame   & Medium  & 2003 Jan 9 03:28:56 & 27.9     & 1.68 \\
MOS1       & Full Frame   & Medium  & 2003 Jan 9 03:06:55 & 29.5     & 0.45 \\
MOS2       & Full Frame   & Medium  & 2003 Jan 9 03:06:55 & 29.5     & 0.46 \\
RGS1       & Spectroscopy & ...     & 2003 Jan 9 03:06:03 & 29.7     & 0.06 \\
RGS2       & Spectroscopy & ...     & 2003 Jan 9 03:06:03 & 29.7     & 0.08 \\
\hline
\end{tabular}
\end{center}
\end{table*}

High resolution X-ray spectra, such as those obtained today with grating spectrometers on board {\it XMM-Newton} and {\it Chandra}, offer the possibility to reconstruct the emission measure distribution ($EMD$) of the emitting plasma, to measure its abundances, and to constrain the electron density $N_{\rm e}$. These diagnostics help to improve our understanding of the X-ray emission from accreting and non accreting young stars. However, to achieve a good $S/N$ ratio in these spectra, bright and nearby sources are needed. The TW~Hydrae association \citep[\object{TWA}, ][and references therein]{ZuckermanWebb2001} is one of the nearest ($\sim55\,{\rm pc}$) and youngest ($\sim10\,{\rm Myr}$) star forming regions and therefore its members are ideal targets for the analysis of X-ray emission from PMS stars by means of high resolution spectroscopy. In the present paper we report on the {\it XMM-Newton} observation of \object{TWA~5} (CD~$-33\degr7795$).
High resolution X-ray spectra of PMS stars have been analyzed in sufficient detail so far for only four other stars: \object{TW~Hydrae} (TW~Hya or TWA~1), \object{HD~98800} (TWA~4), \object{PZ~Tel} and \object{HD~283572}. \object{TW~Hydrae} and \object{HD~98800}, together with \object{TWA~5}, are members of the TWA; \object{PZ~Tel} and  \object{HD~283572} belong to the $\beta$-Pictoris moving group and to the Taurus-Auriga star forming region, respectively.

This paper is organized as follows: in Sect.~\ref{starsample} the principal characteristics of \object{TWA~5} and of the other PMS stars used for comparison are reported; Sect.~\ref{obs} presents the main information about the {\it XMM-Newton} observation of \object{TWA~5} and the methods adopted for the data analysis; in Sect.~\ref{results} we report the results derived, which are discussed and compared with properties of other PMS stars in Sect.~\ref{disc}; we draw our conclusions in Sect.~\ref{concl}.

\section{Star Sample}
\label{starsample}

\object{TWA~5} is a quadruple system located $\sim55$\,pc from the Sun\footnote{Only four \object{TWA} members have measured Hipparcos distances, whose average value, 55\,pc, has been assumed as the distance of \object{TWA~5}.}. The primary, \object{TWA~5A}, is a triple system: a $0\farcs06$ binary resolved by adaptive optics \citep{MacintoshMax2001,BrandekeJayawardhanar2003}, one of the visual components being itself a spectroscopic binary \citep{TorresNeuhauser2001}. All three of them have similar spectral types (M1.5). The secondary, \object{TWA~5B}, is a brown dwarf separated by $2\arcsec$ from the primary \citep{LowranceMcCarthy1999,WebbZuckerman1999}. \object{TWA~5A} does not show any infrared excess indicating no significant amount of circumstellar material \citep{MetchevHillenbrand2004,WeinbergerBecklin2004,UchidaCalvet2004}.  On the other hand, \citet{MohantyJayawardhana2003} measured H$\alpha$ emission typical of accreting PMS stars and signatures of outflows, and concluded that at least one of the components in the \object{TWA~5A} system is a CTTS. It remains currently unclear how the accretion signatures can be reconciled with the lack of evidence of a disk. 
Moreover it is unknown whether the X-ray emitting component of \object{TWA~5A} coincides with the accreting one.

Table~\ref{tab:stars} summarizes the relevant stellar parameters for \object{TWA~5} and the other stars, that we will use for comparison in our study.

\object{TW~Hya} is a single CTTS with enhanced ${\rm H}\alpha$ emission \citep[equivalent width $\sim200$\,\AA, ][]{AlencarBatalha2002,Reid2003} and strong infrared excess \citep{UchidaCalvet2004}. Its X-ray emission, observed with {\it Chandra}/HETGS \citep{KastnerHuenemoerder2002} and {\it XMM-Newton} \citep{StelzerSchmitt2004}, shows peculiar features: the emitting plasma has quite a low temperature ($\log T {\rm (K)} \sim 6.5$), the ${\rm Ne/Fe}$ abundance ratio is as high as a factor 10 in solar photospheric units, and the electron density $N_{\rm e}$, derived from the He-like triplets of \ion{O}{vii} and \ion{Ne}{ix}, is $\sim10^{13}\,{\rm cm^{-3}}$, more than two orders of magnitude above that of typical stellar coronae. Based on these peculiarities it was suggested that X-ray emission from \object{TW~Hya} is produced in an accretion shock rather than in a corona. On the other hand, \citet{Drake2005} has pointed out that the He-like emission line triplets may be affected by photoexcitation due to the UV radiation field. If this were the case the triplet $f/i$ ratio would overestimate the density in the emitting region.

\object{HD~98800} is a WTTS quadruple system, composed by two visual components \object{HD~98800A} and \object{HD~98800B} (whose separation is $0\farcs8$), each of which is a spectroscopic binary. It has been observed with {\it Chandra}/HETGS \citep{KastnerHuenemoerder2004}; from this observation it emerged that its X-ray emission is due mainly to \object{HD~98800A}, and it is produced by plasma at temperatures in the range $\log T {\rm (K)} \sim 6.4-7.0$, with ${\rm Ne/Fe}\sim5$, and low electron density ($N_{\rm e}<10^{12}\,{\rm cm^{-3}}$), typical of stellar coronae.

\object{PZ~Tel} and \object{HD~283572} are two single WTTSs. The X-ray spectrum of \object{PZ~Tel}, gathered with {\it Chandra}/HETGS, has been analyzed by \citet{ArgiroffiDrake2004}. The X-ray spectrum of \object{HD~283572}, observed with both
{\it Chandra}/HETGS and {\it XMM-Newton}, has been studied by \citet{AudardSkinner2005} and by \citet{ScelsiMaggio2005}. For both \object{PZ~Tel} and \object{HD~283572} a typical coronal plasma emerged, with temperatures of $\sim10$\,MK, and with ${\rm Ne/Fe}\sim2-3$ times the solar photospheric ratio.

\section{Observation and Data Analysis}
\label{obs}

\object{TWA~5} was observed for $\sim30$\,ks with {\it XMM-Newton} on 2003 January 9. In Table~\ref{tab:log} we report the observation log for all the instruments.

Both EPIC and RGS data have been processed with SAS V5.4.1 standard tools. We have extracted EPIC source events from a circle with radius of $60''$, centered on the target position. This extraction circle includes 90\% of the source encircled energy. Background events have been extracted from an annular region around the target with inner and outer radii of $60''$ and $90''$. The observation is not affected by significant background contamination due to solar flares and therefore no time screening was required. We have also verified that EPIC spectra are not affected by significant pile-up. The spectral analysis has been performed by adopting the Astrophysical Plasma Emission Database \citep[APED V1.3, ][]{SmithBrickhouse2001}, which assumes ionization equilibrium according to \citet{MazzottaMazzitelli1998}.

\begin{figure}
\resizebox{\hsize}{!}{\includegraphics{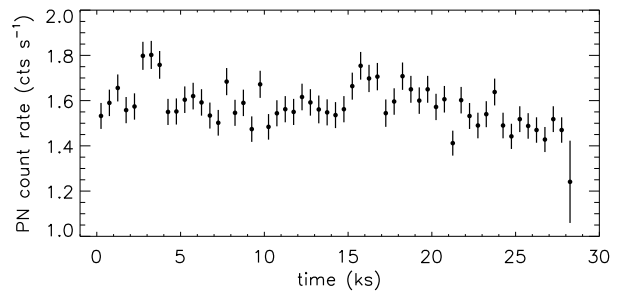}}
\caption{{\it XMM-Newton}/PN light curve of \object{TWA~5} with bin size of 500\,s.}
\label{fig:twa5lightcurve}
\end{figure}

\subsection{EPIC Data Analysis}
\label{EPICanalysis}

\begin{figure*}
\resizebox{0.5\hsize}{!}{\includegraphics{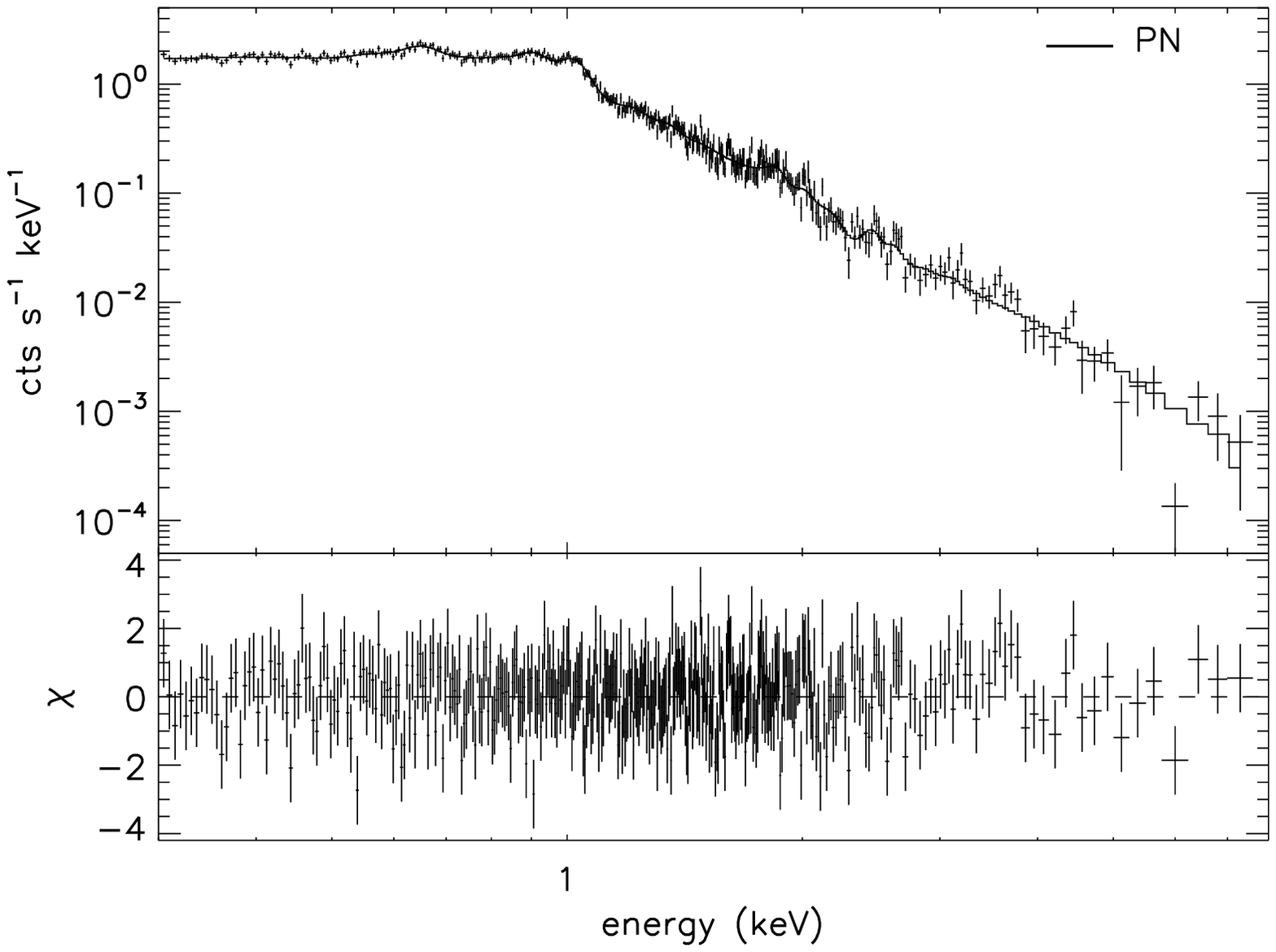}}
\resizebox{0.5\hsize}{!}{\includegraphics{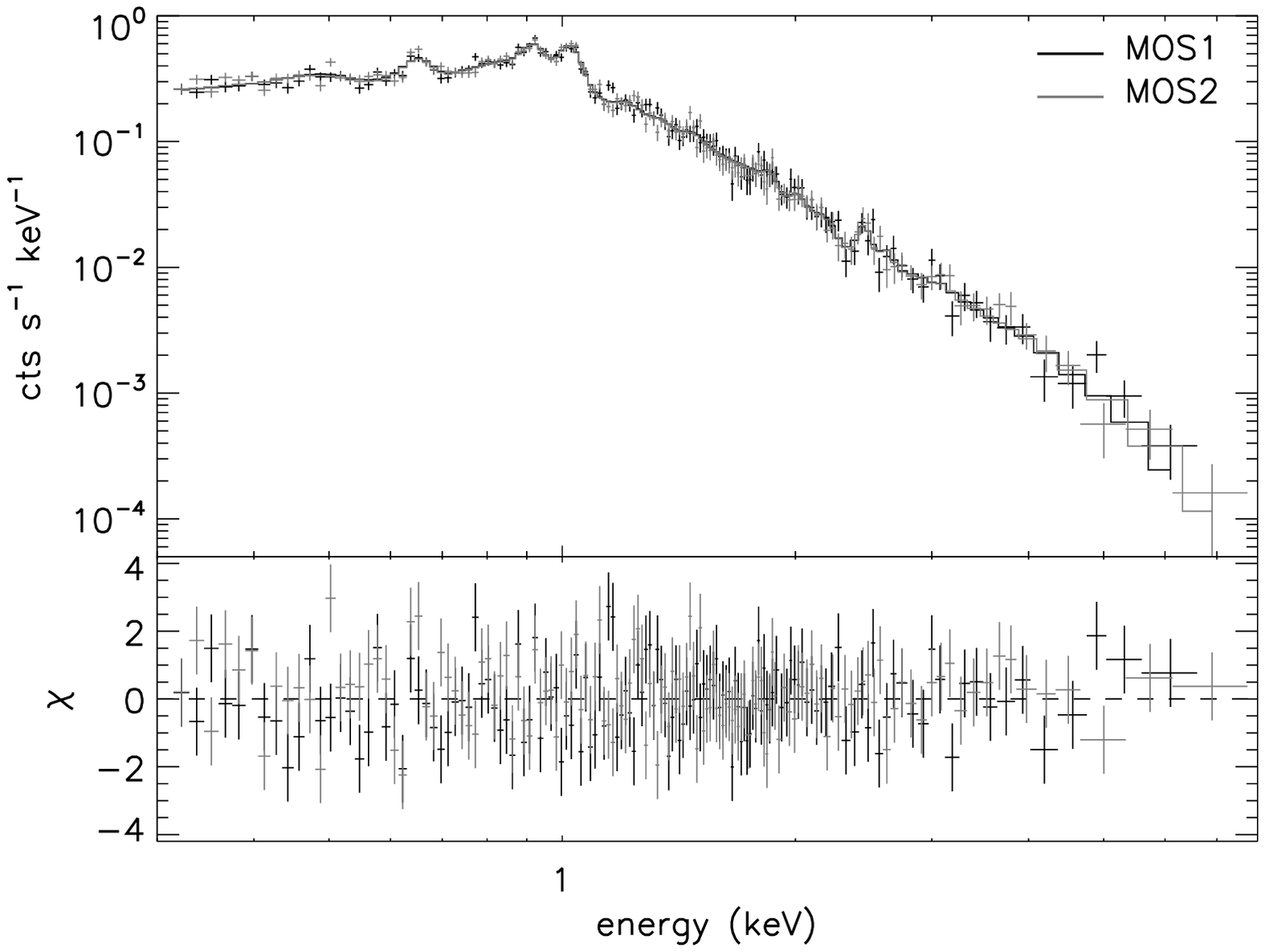}}
\caption{EPIC spectra (PN in upper panel, MOS1 and MOS2 in lower panel) of \object{TWA~5} with best-fit 3-$T$ models superimposed. The lower section of each panel contains residuals. Best-fit parameters are reported in Table~\ref{tab:res}.}
\label{fig:epicspec}
\end{figure*}

The PN light curve of \object{TWA~5} (Fig.~\ref{fig:twa5lightcurve}) does not show strong flare-like events, but an unbinned Kolmogorov-Smirnov test applied to the PN photon arrival times yields a probability of $2\times10^{-7}$ related to the hypothesis of constant emission. This result indicates the presence of significant small amplitude variability. The observed PN and MOS spectra are shown in Fig.~\ref{fig:epicspec}. We have fitted separately the PN and MOS spectra in the energy range $0.3-7.9$\,keV, assuming an absorbed optically-thin plasma model with three thermal components. We have also left as free parameters the abundance of those elements (O, Ne, Mg, Si, S, Fe, and Ni) which significantly improved the fit, while the abundances of the remaining elements (C, N, Al, Ar, and Ca) were tied to the Fe abundance. We have performed the fitting by using XSPEC V11.3.0. The uncertainty on each best-fit parameter, at the 68\% confidence level, has been computed by exploring the $\chi^2$ variation while varying simultaneously all the other free parameters. From the PN best-fit model we have derived  an estimate for the hydrogen column density, $N_{\rm H}\sim3\times10^{20}\,{\rm cm^{-2}}$, and the same value has been found as an upper limit from the analysis of the MOS spectra. This result indicates that the spectra of \object{TWA~5} do not suffer strong absorption. The derived $N_{\rm H}$ value is compatible with that assumed by \citet{JensenCohen1998}, which agrees with the negligible extinction toward the \object{TWA} region. The results obtained from the PN and MOS spectral fitting are reported in Table~\ref{tab:res}. 

\begin{figure*}
\resizebox{1.0\hsize}{!}{\includegraphics{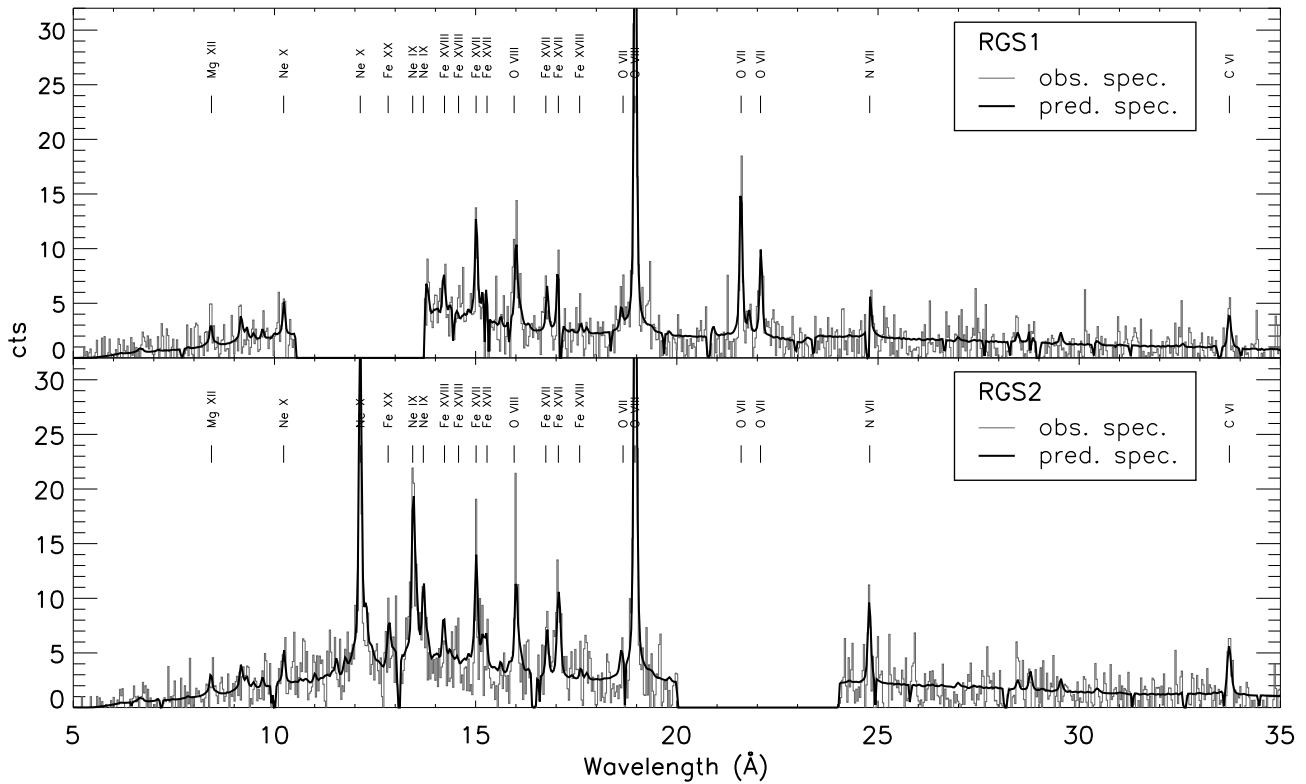}}
\caption{Observed and predicted RGS spectra (rebinned with a wavelength bin of 0.03\,\AA) of \object{TWA~5}.}
\label{fig:rgsspec}
\end{figure*}

\begin{table}[!b]
\begin{center}
\caption{\object{TWA~5} best fit paramenters.}
\label{tab:res}
\begin{tabular}{lccc}
\hline\hline
    & PN                     & MOS                    & RGS                    \\
\hline
 & \multicolumn{3}{c}{Abundances$^{\rm a, b}$ ($A_{X}/A_{X\sun}$)}             \\
\hline
C   & $={\rm Fe}$	           & $={\rm Fe}$	           & $0.26^{+0.30}_{-0.08}$ \\ 
N   & $={\rm Fe}$	           & $={\rm Fe}$	           & $0.55^{+0.61}_{-0.18}$ \\
O   & $0.15^{+0.11}_{-0.07}$ & $0.28^{+0.14}_{-0.14}$ & $0.31^{+0.29}_{-0.07}$ \\
Ne  & $0.34^{+0.27}_{-0.16}$ & $0.93^{+0.37}_{-0.51}$ & $0.92^{+0.74}_{-0.26}$ \\
Mg  & $0.07^{+0.15}_{-0.06}$ & $0.14^{+0.20}_{-0.13}$ & $0.51^{+0.16}_{-0.38}$ \\
Si  & $0.21^{+0.20}_{-0.16}$ & $0.19^{+0.22}_{-0.15}$ & $={\rm Fe}$            \\
S   & $0.32^{+0.36}_{-0.31}$ & $0.36^{+0.38}_{-0.32}$ & $={\rm Fe}$            \\
Fe  & $0.05^{+0.04}_{-0.02}$ & $0.09^{+0.06}_{-0.05}$ & 0.1		                  \\
Ni  & $1.08^{+1.17}_{-0.98}$ & $0.06^{+1.31}_{-0.06}$ & $={\rm Fe}$            \\
\hline
 & \multicolumn{3}{c}{Temperature$^{\rm b}$ (K)}                               \\
\hline
$\log T_{1}$  & $6.17^{+0.56}_{-0.20}$ & $6.56^{+0.14}_{-0.14}$  &             \\
$\log T_{2}$  & $6.70^{+0.17}_{-0.05}$ & $6.90^{+0.19}_{-0.25}$  &             \\
$\log T_{3}$  & $7.27^{+0.09}_{-0.06}$ & $7.27^{+0.33}_{-0.09}$	 &             \\
\hline
 & \multicolumn{3}{c}{Emission Measure$^{\rm b}$ $N_{\rm e}N_{\rm H}V\;({\rm cm^{-3}})$}            \\
\hline
$\log EM_{1}$ & $52.71^{+0.79}_{-0.63}$ & $52.45^{+0.60}_{-0.37}$ &            \\
$\log EM_{2}$ & $53.21^{+0.26}_{-0.31}$ & $52.76^{+0.34}_{-0.70}$ &            \\
$\log EM_{3}$ & $52.72^{+0.10}_{-0.26}$ & $52.64^{+0.15}_{-0.30}$ &            \\
\hline
 & \multicolumn{3}{c}{Column Density$^{\rm b}$ (${\rm 10^{20} cm^{-2}}$)}      \\
\hline
$N_{\rm H}$   & $2.8^{+3.9}_{-2.5}$     & $\le3.3$                & $=1$       \\
\hline
 & \multicolumn{3}{c}{Best Fit Statistics}                                     \\
\hline
$\chi^{2}_{\rm red}$             & 0.89 & 0.99 &     \\
d.o.f.                           & 404  & 285  &     \\
$P(\chi^{2}>\chi^{2}_{\rm obs})$ & 94\% & 54\% &     \\
\hline
\end{tabular}
\end{center}
\begin{list}{}{}
\item[$^{\rm a}$] Solar photospheric abundances are from \citet{AndersGrevesse1989}.
\item[$^{\rm b}$] All the uncertainties correspond to the 68\% confidence level.
\end{list}
\end{table}

\subsection{RGS Data Analysis}
\label{RGSanalysis}

The RGS1 and RGS2 spectra of \object{TWA~5} are shown in Fig.~\ref{fig:rgsspec}. The analysis has been performed using ISIS \citep{HouckDenicola2000} and PINTofALE \citep{KashyapDrake2000}. Our approach is to derive $EMD$ and abundances starting from the line flux measurements. It is known that the line spread function of the RGS spectra is characterized by large wings, making it difficult to identify correctly the continuum level and therefore to measure line fluxes. In order to obtain accurate line flux measurements we have evaluated the continuum level by performing a global fit of the RGS1 and RGS2 spectra. We have adopted a model composed of three isothermal components with variable abundances of C, N, O, Ne, Mg and Fe.  The continuum predicted on the basis of this best-fit model has been used to measure the fluxes of the strongest RGS emission lines. To improve the spectral $S/N$ ratio we have measured the line fluxes by fitting simultaneously RGS1 and RGS2 spectra rebinned with a 0.03\,\AA~wavelength bin. These line fluxes are reported in Table~\ref{tab:lines}\footnote{Table~\ref{tab:lines} is available at the CDS and it contains the following information for each observed line: observed and predicted line wavelength (Cols. 2 and 3), element and ionization state (Col. 4), electronic configurations of the atomic levels (Col. 5), temperature of maximum emissivity (Col. 6), observed line flux (Col. 7).}.

\begin{figure}[t]
\resizebox{\hsize}{!}{\includegraphics{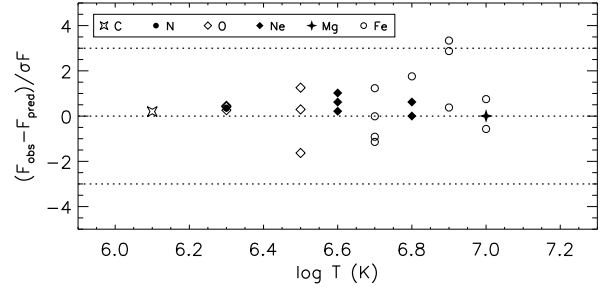}}
\caption{Comparison between observed and predicted line fluxes in the RGS spectra. The predicted values are obtained using the models derived with the MCMC method (see Sect.~\ref{RGSanalysis}).}
\label{fig:checkflux}
\end{figure}

We have reconstructed the $EMD$ and element abundances with the Markov-Chain Monte Carlo (MCMC) method of \citet{KashyapDrake1998} applied to the measured line fluxes. This method performs a search in the $EMD$ and abundances parameter space with the aim of maximizing the probability of obtaining the best match between observed and predicted line fluxes. Some of the main advantages of this method are that it does not need to assume a particular analytical function for the $EMD$, and that it allows to estimate uncertainties on each $EMD$ and abundance value. On the basis of the formation temperature of the selected set of lines, we have adopted a temperature grid ranging from $\log T {\rm (K)} = 6.0$ to $\log T {\rm (K)} = 7.3$, with resolution $\Delta\log T {\rm (K)} = 0.1$, over which to perform the $EMD$ reconstruction. We have assumed a hydrogen column density $N_{\rm H}=10^{20}\,{\rm cm^{-2}}$, compatible with the values derived from the analysis of the EPIC spectra (see Sect.~\ref{EPICanalysis}). In Fig.~\ref{fig:checkflux} we show the comparison between the observed line fluxes and those predicted on the basis of the $EMD$ and abundances derived with the MCMC method.

\begin{table}[!b]
\begin{center}
\caption{Total counts (lines + continuum) in the interval 10-30\,\AA~band vs. Fe abundance.}
\label{tab:feabund}
\begin{tabular}{c@{\hspace{8mm}}ccccc@{\hspace{8mm}}c}
\hline\hline
                                     & \multicolumn{5}{c}{$N_{\rm pred}$} & \multicolumn{1}{c}{$N_{\rm obs}$} \\
${\rm Fe/Fe_{\sun}}^{\rm a}$ & 0.05 & 0.07 & 0.10 & 0.15 & 0.20   &       \\
\hline
RGS1                                 & 2298 & 1841 & 1498 & 1232 & 1099   &  1377 \\
RGS2                                 & 2886 & 2379 & 1998 & 1702 & 1554   &  1874 \\
\hline
\end{tabular}
\end{center}
\begin{list}{}{}
\item[$^{\rm a}$] Abundance referred to the solar photospheric value from \citet{AndersGrevesse1989}.
\end{list}
\end{table}

Since line fluxes depend on the product of the $EMD$ with the element abundances, the adopted method provides the $EMD$ scaled by the Fe abundance, and the abundance ratio of each element with respect to Fe. However, it is worth noting that the continuum emission,  depends strongly on the amount of emission measure, and weakly on the absolute abundances of elements heavier than He. In fact the continuum emission is due to three processes: bremsstrahlung radiation, radiative recombination and two-photons emission. For the temperatures involved in the plasma of \object{TWA~5} the main contribution to the continuum is due to bremsstrahlung radiation which depends very weakly on the heavy element abundances. Therefore, after performing the MCMC reconstruction, we have considered several models assuming different absolute Fe abundances and therefore different global scaling factors for the $EMD$ distribution. For each of these models we have compared the predicted and observed continuum levels. Since
it is hard to identify correctly the continuum level in RGS spectra, as already mentioned above, we have also compared the observed and predicted total emission (spectral lines + continuum) as a further check. In Table~\ref{tab:feabund} we report the explored Fe abundances, and the corresponding total counts for the simulated spectra $N_{\rm pred}$, to be compared with the observed total number of counts, $N_{\rm obs}$. With this procedure we have determined the absolute Fe abundance, and therefore the absolute position of the $EMD$ and the absolute abundances of all the other elements. The resulting Fe abundance is $0.1$ times the solar photospheric value of \cite{AndersGrevesse1989}, with an uncertainty smaller than a factor 2. As a final cross-check, we have verified that the predicted continuum level agrees with the continuum used for the line flux measurements. The abundances resulting from the RGS analysis are reported in Table~\ref{tab:res}.

\section{Results}
\label{results}

The X-ray luminosity of \object{TWA~5}, computed in the interval $6-20$\,\AA~from the best-fit models of PN, MOS and RGS spectra, is 8.3, 6.7 and $6.7\times10^{29}\,{\rm erg\,s^{-1}}$, respectively. The derived luminosities are compatible within the best-fit parameter errors. As shown by \citet{TsuboiMaeda2003}, from a {\it Chandra}/ACIS-S observation, the X-ray emission is essentially due to the primary \object{TWA~5A}. In fact the {\it Chandra} observation was able to resolve the brown dwarf \object{TWA~5B} from the primary \object{TWA~5A}, and \citeauthor{TsuboiMaeda2003} measured for \object{TWA~5B} an X-ray luminosity of $4\times10^{27}\,{\rm erg\,s^{-1}}$.

\begin{figure}
\resizebox{\hsize}{!}{\includegraphics{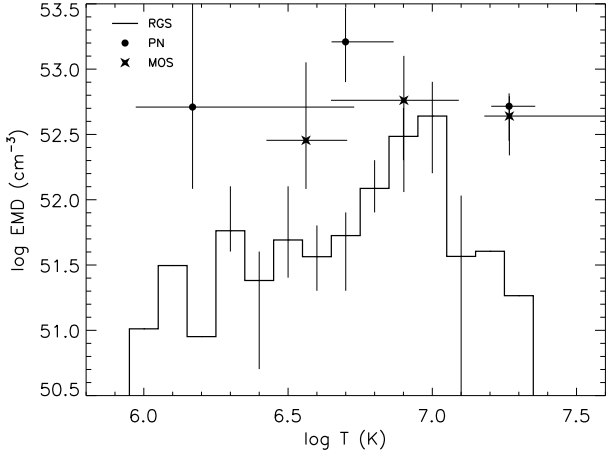}}
\caption{Emission measure distribution of \object{TWA~5} obtained from the RGS line fluxes with the MCMC method, along with the values obtained with the 3-$T$ best-fit model of PN and MOS spectra. The uncertainties correspond to 68\% statistical confidence level.}
\label{fig:twa5emd}
\end{figure}

\subsection{Emission Measure Distribution}

In Fig.~\ref{fig:twa5emd} we report the $EMD$ vs. temperature derived from the EPIC and RGS spectral analysis of \object{TWA~5}\footnote{We note that the MCMC method explores preferentially $EMD$ bins which are best constrained by the selected emission lines. Since error estimation depends on the quality of the sampling, statistical uncertainties are estimated only for those $EMD$ bins explored many times \citep{KashyapDrake1998}.}. All the instruments detect the strongest thermal component at $\log T {\rm (K)} \sim 6.7-7.0$, but the RGS spectra are not able to probe the hottest plasma component at $\log T {\rm (K)} \sim 7.3$, detected by EPIC.  The reason of this result is the different effective area of EPIC and RGS in the hardest part of the X-ray spectra ($E \ga 2$\,keV). In principle, the high-temperature tail could be probed by exploiting a number of \ion{Fe}{xxii-xxiv} lines, which fall in the wavelength region $10.7-11.8$\,\AA~ covered by RGS, but the emissivity of these lines is relatively low and the RGS resolution too poor for this purpose.

\subsection{Abundances}
\label{resabund}

\begin{figure}
\resizebox{\hsize}{!}{\includegraphics{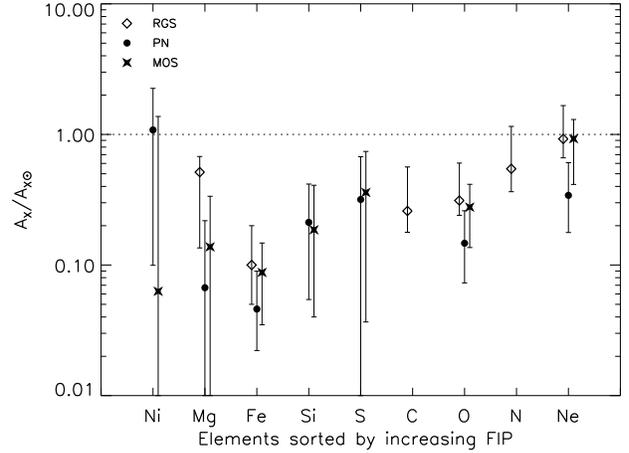}}
\caption{Abundances of \object{TWA~5}, with respect to solar photospheric values \citep{AndersGrevesse1989}, obtained using the RGS line fluxes with the MCMC method, along with the values obtained with the 3-$T$ best-fit model of PN and MOS spectra. The uncertainties correspond to the 68\% confidence level, except the error bar of Fe derived from RGS data, which has been obtained with a different procedure, as discussed in Sects.~\ref{RGSanalysis} and \ref{resabund}.}
\label{fig:twa5abund}
\end{figure}

In Fig.~\ref{fig:twa5abund} (and in Table~\ref{tab:res}) we show the element abundances, in solar photospheric units \citep{AndersGrevesse1989}, derived from the spectra obtained with each instrument. The elements are sorted along the abscissa by increasing values of first ionization potential (FIP). The abundances of C and N, which have their strongest emission lines in the low-energy part of the observed spectral range ($\lambda\sim25-35$\,\AA, or $E\sim0.35-0.5$\,keV), are derived only from the RGS. On the other hand, the abundances of Si and S are estimated only from the EPIC spectra since their H-like and He-like lines fall at high energies, and they cannot be constrained by the RGS. Note that the Fe abundance derived from RGS data has been estimated with a procedure (Sect.~\ref{RGSanalysis}) which does not allow to obtain a formal statistical uncertainty, however we are confident that it cannot be off by more than a factor 2, as explained in Sect.~\ref{RGSanalysis}. We note that the abundances derived from different instruments are compatible within statistical uncertainties. However some systematic differences come out, as briefly discussed in Sect.~\ref{modelcomp}. From the derived abundances it emerges that the X-ray emitting plasma of \object{TWA~5} is metal depleted.

\subsection{Electron Density}

\begin{figure}
\resizebox{\hsize}{!}{\includegraphics{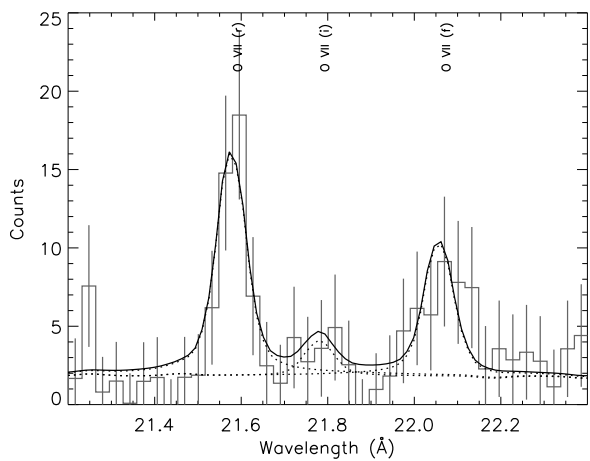}}
\caption{\ion{O}{vii} triplet in the RGS1 spectrum of \object{TWA~5}, rebinned with a wavelength bin of 0.03\,\AA, with best-fit line profiles: individual lines ({\it dotted line}) and their sum ({\it solid line}).}
\label{fig:twa5OVII}
\end{figure}

We have evaluated the plasma electron density from the analysis of the \ion{O}{vii} He-like triplet. The other He-like triplets which fall in the RGS spectral range were either too weak (\ion{N}{vi}, \ion{Mg}{xi}, \ion{Si}{xiii}), or heavily blended with other strong lines (\ion{Ne}{ix}) to be analyzed. In Fig.~\ref{fig:twa5OVII} we show the RGS1 spectrum (rebinned with a 0.03\,\AA~bin size) in the \ion{O}{vii} triplet region with superimposed the best-fit curves. The measured ratio of forbidden $f$ and intercombination $i$ line fluxes is $3.8\pm2.5$ (see Table \ref{tab:lines} for line fluxes), which yields an upper limit of $\sim10^{11}\,{\rm cm^{-3}}$ for $N_{\rm e}$, adopting the predicted $f/i$ ratios of \citet{SmithBrickhouse2001}.

\subsection{Comparison between Different Models}
\label{modelcomp}

We have performed separate analyses of the PN, MOS and RGS spectra for several reasons. The main reason is that an approach based on fluxes of selected individual lines, measurable only in the RGS data, provide us with the most reliable results for the element abundances and for the plasma $EMD$. Moreover, independent analyses of EPIC data offer the opportunity to compare the results of the different {\it XMM-Newton} instruments. These comparisons allow us to investigate the robustness of each measurement, and therefore they are useful to test the reliability of results based on EPIC data only, in the broader contest of observations of X-ray coronal sources with no high resolution spectrum available.

As already mentioned in Sect.~\ref{resabund}, abundance estimation from different {\it XMM-Newton} instruments turns out to be quite robust, at least within the statistical uncertainties of a typical {\it XMM-Newton} observation ($\sim30$\,ks exposure time, in the present case). However, the abundances of Mg, Fe, O, and Ne obtained from the PN fitting are systematically lower than the corresponding values based on the analysis of the MOS and RGS spectra, which agree among themselves. On the other hand, the emission measure values derived from the PN analysis are larger than those obtained from the MOS and RGS spectra, and the X-ray luminosity predicted by the PN model is $\sim 20$\% higher than in the other two cases. Although, all the differences are within the statistical uncertainties, it is conceivable that the higher spectral resolution of the MOS detector, with respect to the EPIC/PN, allows to disentangle better the contributions of lines and continuum, and therefore to constrain the absolute values of $EMD$ and abundances. Moreover, EPIC/MOS and RGS share the same X-ray telescopes and hence their cross-calibration is better determined, while residual calibration problems of the PN instrument may cause the differences in the fitting results described above.

EPIC models are able to provide a good global description of the source plasma, but limited to models with few free parameters, while RGS spectra allow to derive a more detailed model but -- with the available signal to noise ratio -- they fail to detect plasma components with temperatures higher than $T\ge10$\,MK, due to the smaller energy range covered by RGS with respect to EPIC.

In order to cross-check the three models we have compared each of them with the spectra of different {\it XMM-Newton} instruments and we have computed reduced $\chi^2$ values. The best-fit 3-$T$ models of PN and MOS globally describe RGS spectra reasonably well ($\chi^{2}_{\rm red}=1.8$ and $\chi^{2}_{\rm red}=1.3$ respectively, with 817 d.o.f.), but not as well as the line-based $EMD$ model\footnote{This is not obvious since our RGS $EMD$ model has been obtained from the analysis of selected line fluxes and not from a global spectral fitting.} ($\chi^{2}_{\rm red}=1.0$ with 817 d.o.f.). As already noticed the RGS model misses the higher temperature components and hence it underestimates the PN and MOS spectra at high energy. Finally, both RGS and MOS models show some disagreement with the PN spectrum in the low energy range.

\section{Discussion}
\label{disc}

In this section we discuss the results obtained for \object{TWA~5}, in terms of $EMD$, abundances and density, and compare them with those for the CTTS \object{TW~Hya}, and the other WTTSs in our sample (Sect.~\ref{starsample}). We stress that all results are based on high resolution X-ray spectra. It must be recalled that \object{TWA~5A} is a triple system, and so far we have not been able to determine whether the X-ray emission and the accretion signatures emerge from the same star.

The analysis of both the EPIC and RGS data has shown that the X-ray emission of \object{TWA~5} is mainly produced by hot plasma ($T\sim 10$\,MK). The analysis of the \ion{O}{vii} triplet has indicated a typical coronal electron density ($N_{\rm e}\le10^{11}\,{\rm cm^{-3}}$). These characteristics are similar to those found in WTTSs \citep{KastnerHuenemoerder2004,ArgiroffiDrake2004,ScelsiMaggio2005} and magnetically active late-type main sequence stars when high resolution X-ray spectroscopy is used \citep[see e.g. ][]{NessGudel2004}. Peculiar features of \object{TWA~5} are its
very low metallicity (${\rm Fe/Fe_{\sun}}\sim 0.1$) and its extremely high abundance ratio ${\rm Ne/Fe}\sim 10$. Such high values for the ${\rm Ne/Fe}$ have been observed only in a few very active stars (\object{HR~1099}, \object{UX~Ari}, \object{II~Peg}) having average coronal temperatures larger than those of \object{TWA~5} (see below), and it has been ascribed to the so-called inverse FIP effect (see discussion below). On the other hand, a low metallicity (${\rm Fe/Fe_{\sun}}\sim 0.2$) and the same Ne/Fe ratio have been measured for \object{TW~Hya}, the only unambiguous CTTS studied so far at high spectral resolution in X-rays. Hence, it is an interesting issue why \object{TWA~5} shares the same chemical peculiarities with \object{TW~Hya}, in spite of having other thermal characteristics.

\begin{figure}
\resizebox{\hsize}{!}{\includegraphics{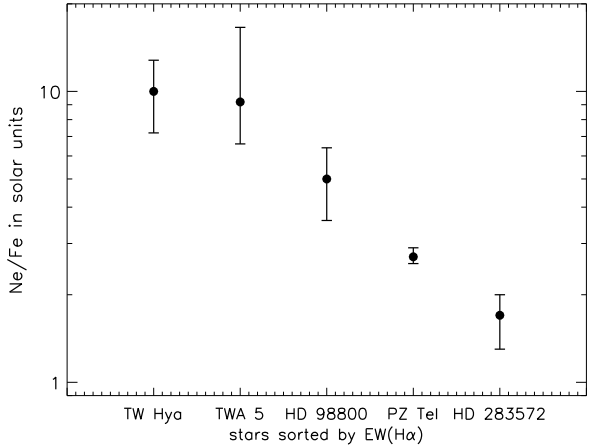}}
\caption{Ne/Fe abundance ratio in solar photospheric units \citep{AndersGrevesse1989} for the PMS stars listed in Table~\ref{tab:stars} and sorted by H$\alpha$ equivalent width. We have assumed that the abundance uncertainties of \object{HD~98800} are equal to those of \object{TW~Hya}, because the observed Chandra spectra of these two stars have similar $S/N$.}
\label{fig:netofe1}
\end{figure}

We recall that \object{TW~Hya} presents spectral characteristics compatible with a model of X-ray emission driven, or at least affected, by the infalling accretion stream \citep{KastnerHuenemoerder2002,StelzerSchmitt2004}. In fact, all the X-ray properties of \object{TW~Hya} (e.g. its low plasma temperature, high density, and metal depletion) suggest that the emitting plasma forms in the shock region where the infall streams reach the stellar surface. Among the peculiar characteristics of \object{TW~Hya} the very low abundances of all the metals in the emitting plasma appear to be compatible with the accretion scenario. In fact, \citeauthor{StelzerSchmitt2004} proposed that Fe and other heavy elements in the accretion disk condense into dust grains \citep[see e.g. ][]{SavageSembach1996} which possibly settle into the disk midplane, while other elements like N, remain in the gas phase. Neon and other noble elements, should also not remain locked onto dust grains, but rather be part of the gas phase \citep{FrischSlavin2003}. Since the accreting material is largely composed by gas rather than dust \citep{TakeuchiLin2002}, the accreting stream is expected to display a high Ne/Fe abundance ratio. This material falls onto the stellar surface and there, heated to temperatures of few MK by the ensuing shock, produces X-ray radiation revealing its anomalous chemical composition.  The intriguing point is that \object{TWA~5} presents exactly the same abundance ratios of \object{TW~Hya}, and in particular a ${\rm Ne/Fe}\sim10$, but lacks all the other indications for accretion-related X-ray emission. In Fig.~\ref{fig:netofe1} we show the Ne/Fe ratio for the PMS stars in our sample. The stars are sorted along the abscissa by increasing value of H$\alpha$ equivalent width, in order to separate the accreting CTTSs, on the left part of the diagram, from the non accreting WTTSs. This plot suggests that CTTSs tend to have Ne/Fe higher than WTTSs. 

As already hinted above the Ne/Fe ratio could be influenced also by FIP-related effects: in the solar corona, and in particular in long-lived active regions, and in late type stars with low activity levels, abundances of elements with low FIP appear to be enhanced with respect to the high FIP elements \citep[see ][ and references therein]{FeldmanWiding2003}, using photospheric abundances as a reference. On the other hand, more active stars present an overabundance of high FIP elements with respect to low FIP elements, the so called inverse FIP effect \citep{BrinkmanBehar2001,DrakeBrickhouse2001,AudardGudel2003}. Early models to explain the FIP effect involve ion-neutral fractionation in the chromosphere \citep{Geiss1982,Meyer1996}, but they do not provide a satisfactory explanation for the selective enhancement of some elements in the corona \citep[see][ for a recent review]{Gudel2004}. Most recently, \citet{Laming2004} has proposed a new model which tries to explain both a FIP and an inverse FIP effect as a result of the pondermotive forces related to chromospheric Alfv\'en waves acting on ions of different species. The Ne/Fe ratio is a good indicator of the coronal abundance pattern since Ne has a high FIP value (21.6\,eV), while Fe is a low FIP element (7.9\,eV), and strong lines from both elements have close wavelengths at similar coronal temperatures. Stars with high activity level usually show ${\rm Ne/Fe}\sim1-5$, and only few active binaries present ${\rm Ne/Fe}\sim10$ \citep{BrinkmanBehar2001,DrakeBrickhouse2001,HuenemoerderCanizares2001,AudardGudel2003}.  \citet{Gudel2004} shows that the Ne/Fe ratio tends to increase for increasing average coronal temperature, with the above extreme value reached by stars with $\overline{T}_{c}\sim15$\,MK. For comparison, \object{TWA~5} has $\overline{T}_{c}\approx9$\,MK and stars of comparable temperature in the sample studied by \citeauthor{Gudel2004} show Ne/Fe in the range $1-5$.

\begin{figure}
\resizebox{\hsize}{!}{\includegraphics{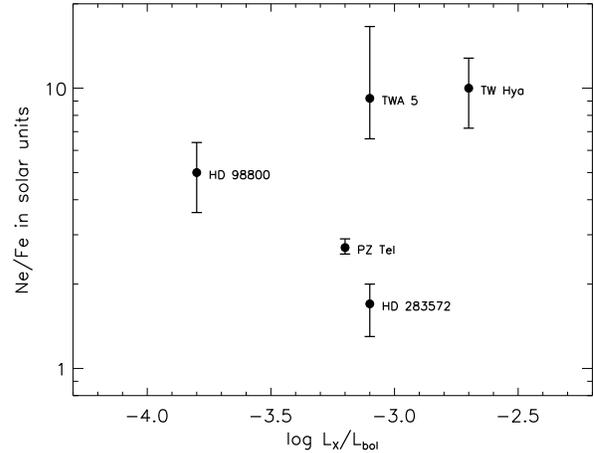}}
\caption{Ne/Fe abundance ratio in solar photospheric units \citep{AndersGrevesse1989}, for the PMS stars listed in Table~\ref{tab:stars}, vs. the X-ray to bolometric luminosity ratio. We have assumed that the abundance uncertainties of \object{HD~98800} are equal to those of \object{TW~Hya}, because the observed Chandra spectra of these two stars have similar $S/N$.}
\label{fig:netofe2}
\end{figure}

To explore further this inverse FIP effect scenario we have plotted the Ne/Fe ratio for the PMS star sample vs. $L_{\rm X}/L_{\rm bol}$ in Fig.~\ref{fig:netofe2}. In fact active stars do show a correlation between coronal abundances and the activity level \citep{SinghDrake1999,GudelAudard2002,AudardGudel2003}. If the differences in Ne/Fe ratio among the stars in our sample were caused by a similar FIP-related effect we would expect to see a correlation between Ne/Fe and $L_{\rm X}/L_{\rm bol}$. For comparison purposes, we have included in the plot also \object{TW~Hya}, even if its X-ray emission is likely not due to coronal activity. This plot does not show any clear trend, even if we do not consider \object{TW~Hya}. This result might be due to the small number of PMS stars studied so far with high resolution X-ray spectroscopy, and to the fact that most of these stars are in the saturated emission regime where $L_{\rm X}/L_{\rm bol}\sim10^{-3}$. If we insist that an inverse FIP effect is responsible for the observed Ne/Fe ratio of \object{TWA~5}, it still remains unclear why stars with similar characteristics (age, plasma temperature, $L_{\rm X}/L_{\rm bol}$) do show Ne/Fe values which differ by about a factor 10, as in the case of \object{TWA~5}, \object{PZ~Tel} and \object{HD~283572}. Hence we argue that \object{TWA~5}, and even more clearly \object{TW~Hya}, appear to be outliers with respect to other active stars.

In conclusion we can tentatively depict three different scenarios in order to interpret the characteristics of the X-ray emitting plasma in \object{TWA~5}.

Since \object{TWA~5} appears to contain a CTTS \citep{MohantyJayawardhana2003}, the high Ne/Fe might be due to an accretion process, as already suggested in the case of \object{TW~Hya}. In this scenario, the X-ray emission from \object{TWA~5} should be produced by shock heated plasma at the base of the accretion column. However, shock temperatures are expected to be lower than the values derived from the X-ray spectrum of \object{TWA~5}, and this occurrence is not in favor of accretion-related X-ray emission. This scenario is also questioned by the analysis of X-ray emission from CTTSs and WTTSs in the L1551 region, discussed by \citet{FavataGiardino2003}: they derive ${\rm Ne/Fe}\sim4$ for the three WTTSs and no indication of high Ne/Fe for the two CTTSs in their stellar sample\footnote{However, the results on L1551 region are based on low resolution EPIC spectra, and therefore they may not be directly compared to ours.}. Most recently the analysis of the {\it XMM-Newton}/PN spectrum of the CTTS \object{BP~Tau} revealed a hot plasma, while the \ion{O}{vii} lines suggested a high electron density \citep{SchmittRobrade2005}. These results on \object{BP~Tau} indicate that shock heated and coronal plasma may be both present in CTTSs.

The second scenario is based on the consideration that \object{TWA~5} has $\log (L_{X}/L_{\rm bol}) \sim-3$, at the saturation level for active stars. Therefore, the high Ne/Fe ratio may be related to the same mechanism which produces the inverse FIP effect in the coronae of other active stars.  Under this hypothesis the accretion process does not play a major role in the X-ray emission of \object{TWA~5}, which is instead produced by magnetically confined hot plasma. However the Ne/Fe ratio of \object{TWA~5} appears to be to high by a factor $2-5$ with respect to stars with similar average coronal temperature.

Finally, we note that both \object{TW~Hya} and \object{TWA~5} belong to the same young association, and share the same value of Ne/Fe. Therefore, the third hypothesis is that their anomalous abundances originate from the molecular cloud from which the two stars formed. Such a scenario, in which the measured abundances are related to those of the primordial material implies that the molecular cloud was Fe depleted. In order to confirm or reject this hypothesis the abundances of other members of the \object{TWA} need to be determined. Note that in the case of \object{HD~98800}, a member of \object{TWA}, the ratio ${\rm Ne/Fe}\sim5$ was derived by \citet{KastnerHuenemoerder2004} from a spectrum affected by low S/N ratio which did not allow these authors to perform a detailed $EMD$ analysis. As a consequence the derived Ne/Fe ratio is uncertain since it depends strongly on the $EMD$ shape.

\section{Conclusions}
\label{concl}

We have analyzed the EPIC and RGS data of the CTTS \object{TWA~5} inferring the emitting plasma characteristics: the X-ray emission reveals a hot plasma ($T \sim 10^{7}$\,K) with low electron density ($N_{\rm e}\le 10^{11}\,{\rm cm^{-3}}$) and low metallicity (${\rm Fe/Fe_{\sun}}\sim0.1$). These findings suggest that X-rays may be generated by magnetically-confined coronal plasma strongly influenced by an inverse FIP effect. However stars with coronal temperatures comparable with that of \object{TWA~5} show lower Ne/Fe ratios \citep{Gudel2004}. The ${\rm Ne/Fe}\sim10$ abundance ratio measured for \object{TWA~5} leaves open the issue of the X-ray production mechanism, since the same Ne/Fe has been measured for the CTTS \object{TW~Hya}, where this result has been interpreted as evidence for the shock heated accreting material as responsible for the X-ray emission. An alternative explanation we propose is that the peculiar abundance ratio could be a characteristics of the primeval gas from which all members of the \object{TWA} formed.

\begin{acknowledgements}
CA, AM, GP and BS acknowledge partial support for this work by Agenzia Spaziale Italiana and by Ministero dell'Istruzione, Universit\`a e Ricerca. Based on observations, GTO data by MPE from PI B. Aschenbach, obtained with {\it XMM-Newton}, an ESA science mission with instruments and contributions directly funded by ESA Member States and NASA.
\end{acknowledgements}

\bibliographystyle{aa}
\bibliography{twa5}

\begin{longtable}{lrrllcr@{ \hspace{2mm} $\pm$ \hspace{2mm} }l}
\caption{Strongest RGS lines of TWA~5.} \\
\label{tab:lines} \\
\hline\hline
 \multicolumn{1}{c}{Label} & \multicolumn{1}{c}{$\lambda_{\rm obs}^{\rm a}$} & \multicolumn{1}{c}{$\lambda_{\rm pred}^{\rm a}$} & \multicolumn{1}{c}{Ion} & \multicolumn{1}{c}{Transition}                  & $\log T_{\rm max}^{\rm b}$ & $(F$ & $\sigma F\;)^{\rm c}$ \\
                           & \multicolumn{1}{c}{(\AA)}                       & \multicolumn{1}{c}{(\AA)}                        &                         & \multicolumn{1}{c}{$(upper \rightarrow lower)$} & (K)                        & \multicolumn{2}{c}{${\rm(10^{-6}\,ph\;s^{-1}\,cm^{-2})}$} \\
\hline
\endfirsthead
\caption{continued.} \\
\hline\hline
 \multicolumn{1}{c}{Label} & \multicolumn{1}{c}{$\lambda_{\rm obs}^{\rm a}$} & \multicolumn{1}{c}{$\lambda_{\rm pred}^{\rm a}$} & \multicolumn{1}{c}{Ion} & \multicolumn{1}{c}{Transition}                  & $\log T_{\rm max}^{\rm b}$ & $(F$ & $\sigma F\;)^{\rm c}$ \\
                           & \multicolumn{1}{c}{(\AA)}                       & \multicolumn{1}{c}{(\AA)}                        &                         & \multicolumn{1}{c}{$(upper \rightarrow lower)$} & (K)                        & \multicolumn{2}{c}{${\rm(10^{-6}\,ph\;s^{-1}\,cm^{-2})}$} \\
\hline
\endhead
\hline
\endfoot
\hline
\multicolumn{8}{p{1.0\textwidth}}{
\begin{list}{}{}
\item[$^{\rm a}$]~Observed and predicted (APED database) wavelengths. In the cases of unresolved blends, identified by the same label number, we list the main components in order of increasing predicted wavelength.
\item[$^{\rm b}$]~Temperature of maximum emissivity.
\item[$^{\rm c}$]~Line fluxes with uncertainties at the 68\% confidence level obtained by fitting simultaneously RGS1 and RGS2 spectra. In the cases of unresolved blends, identified by the same label number, we report only the total flux of the blended lines.
\end{list}
}
\endlastfoot
  1a &              8.43 &            8.4192 & \ion{Mg}{    xii} &                                      $2p~^2P_{3/2}\; \rightarrow \;1s~^2S_{1/2}$ &   7.00 &                                        10.9 &  3.9 \\
  1b & $\cdot\cdot\cdot$ &            8.4246 & \ion{Mg}{    xii} &                                      $2p~^2P_{1/2}\; \rightarrow \;1s~^2S_{1/2}$ &   7.00 &              \multicolumn{2}{c}{$\cdot\cdot\cdot$} \\
  2a &             10.23 &           10.2385 & \ion{Ne}{      x} &                                      $3p~^2P_{3/2}\; \rightarrow \;1s~^2S_{1/2}$ &   6.80 &                                        11.5 &  3.1 \\
  2b & $\cdot\cdot\cdot$ &           10.2396 & \ion{Ne}{      x} &                                      $3p~^2P_{1/2}\; \rightarrow \;1s~^2S_{1/2}$ &   6.80 &              \multicolumn{2}{c}{$\cdot\cdot\cdot$} \\
  3a &             12.13 &           12.1240 & \ion{Fe}{   xvii} &                       $2s^22p^5(^2P)4d~^1P_{1}\; \rightarrow \;2s^22p^6~^1S_{0}$ &   6.80 &                                        74.7 &  8.2 \\
  3b & $\cdot\cdot\cdot$ &           12.1321 & \ion{Ne}{      x} &                                      $2p~^2P_{3/2}\; \rightarrow \;1s~^2S_{1/2}$ &   6.80 &              \multicolumn{2}{c}{$\cdot\cdot\cdot$} \\
  3c & $\cdot\cdot\cdot$ &           12.1375 & \ion{Ne}{      x} &                                      $2p~^2P_{1/2}\; \rightarrow \;1s~^2S_{1/2}$ &   6.80 &              \multicolumn{2}{c}{$\cdot\cdot\cdot$} \\
  3d & $\cdot\cdot\cdot$ &           12.1610 & \ion{Fe}{  xxiii} &                              $1s^22s3s~^1S_{0}\; \rightarrow \;1s^22s2p~^1P_{1}$ &   7.20 &              \multicolumn{2}{c}{$\cdot\cdot\cdot$} \\
  4a &             12.31 &           12.2660 & \ion{Fe}{   xvii} &                       $2s^22p^5(^2P)4d~^3D_{1}\; \rightarrow \;2s^22p^6~^1S_{0}$ &   6.80 &                                         6.8 &  4.2 \\
  4b & $\cdot\cdot\cdot$ &           12.2840 & \ion{Fe}{    xxi} &                      $1s^22s^22p3d~^3D_{1}\; \rightarrow \;1s^22s^22p^2~^3P_{0}$ &   7.00 &              \multicolumn{2}{c}{$\cdot\cdot\cdot$} \\
  5a &             12.82 &           12.8240 & \ion{Fe}{     xx} &            $1s^22s^22p_{1/2}2p_{3/2}3d_{3/2}\; \rightarrow \;2s^22p^3~^4S_{3/2}$ &   7.00 &                                        13.0 &  3.9 \\
  5b & $\cdot\cdot\cdot$ &           12.8460 & \ion{Fe}{     xx} &            $1s^22s^22p_{1/2}2p_{3/2}3d_{3/2}\; \rightarrow \;2s^22p^3~^4S_{3/2}$ &   7.00 &              \multicolumn{2}{c}{$\cdot\cdot\cdot$} \\
  5c & $\cdot\cdot\cdot$ &           12.8640 & \ion{Fe}{     xx} &            $1s^22s^22p_{1/2}2p_{3/2}3d_{5/2}\; \rightarrow \;2s^22p^3~^4S_{3/2}$ &   7.00 &              \multicolumn{2}{c}{$\cdot\cdot\cdot$} \\
  6a &             13.44 &           13.4473 & \ion{Ne}{     ix} &                                      $1s2p~^1P_{1}\; \rightarrow \;1s^2~^1S_{0}$ &   6.60 &                                        37.9 &  7.9 \\
  6b & $\cdot\cdot\cdot$ &           13.4620 & \ion{Fe}{    xix} &                       $2s^22p^3(^2D)3d~^3S_{1}\; \rightarrow \;2s^22p^4~^3P_{2}$ &   6.90 &              \multicolumn{2}{c}{$\cdot\cdot\cdot$} \\
  7a &             13.54 &           13.4970 & \ion{Fe}{    xix} &            $1s^22s^22p_{1/2}2p_{3/2}^23d_{3/2}\; \rightarrow \;2s^22p^4~^3P_{2}$ &   6.90 &                                        17.0 &  6.8 \\
  7b & $\cdot\cdot\cdot$ &           13.5070 & \ion{Fe}{    xxi} &                          $1s^22s2p_{1/2}^23s\; \rightarrow \;1s^22s2p^3~^3D_{1}$ &   7.00 &              \multicolumn{2}{c}{$\cdot\cdot\cdot$} \\
  7c & $\cdot\cdot\cdot$ &           13.5180 & \ion{Fe}{    xix} &                       $2s^22p^3(^2D)3d~^3D_{3}\; \rightarrow \;2s^22p^4~^3P_{2}$ &   6.90 &              \multicolumn{2}{c}{$\cdot\cdot\cdot$} \\
  7d & $\cdot\cdot\cdot$ &           13.5531 & \ion{Ne}{     ix} &                                      $1s2p~^3P_{1}\; \rightarrow \;1s^2~^1S_{0}$ &   6.60 &              \multicolumn{2}{c}{$\cdot\cdot\cdot$} \\
  8a &             13.70 &           13.6450 & \ion{Fe}{    xix} &                       $2s^22p^3(^2D)3d~^3F_{3}\; \rightarrow \;2s^22p^4~^3P_{2}$ &   6.90 &                                        16.4 &  4.6 \\
  8b & $\cdot\cdot\cdot$ &           13.6990 & \ion{Ne}{     ix} &                                      $1s2s~^3S_{1}\; \rightarrow \;1s^2~^1S_{0}$ &   6.60 &              \multicolumn{2}{c}{$\cdot\cdot\cdot$} \\
  8c & $\cdot\cdot\cdot$ &           13.7458 & \ion{Fe}{    xix} &                       $2s^22p^3(^2D)3d~^1F_{3}\; \rightarrow \;2s^22p^4~^1D_{2}$ &   6.90 &              \multicolumn{2}{c}{$\cdot\cdot\cdot$} \\
  9a &             14.23 &           14.2080 & \ion{Fe}{  xviii} &          $1s^22s^22p_{1/2}2p_{3/2}^33d_{5/2}\; \rightarrow \;2s^22p^5~^2P_{3/2}$ &   6.90 &                                        12.2 &  2.6 \\
  9b & $\cdot\cdot\cdot$ &           14.2080 & \ion{Fe}{  xviii} &                   $2s^22p^4(^1D)3d~^2D_{5/2}\; \rightarrow \;2s^22p^5~^2P_{3/2}$ &   6.90 &              \multicolumn{2}{c}{$\cdot\cdot\cdot$} \\
  9c & $\cdot\cdot\cdot$ &           14.2560 & \ion{Fe}{  xviii} &          $1s^22s^22p_{1/2}2p_{3/2}^33d_{5/2}\; \rightarrow \;2s^22p^5~^2P_{3/2}$ &   6.90 &              \multicolumn{2}{c}{$\cdot\cdot\cdot$} \\
 10a &             14.58 &           14.4856 & \ion{Fe}{  xviii} &                   $2s^22p^4(^1D)3d~^2S_{1/2}\; \rightarrow \;2s^22p^5~^2P_{1/2}$ &   6.90 &                                        10.5 &  2.5 \\
 10b & $\cdot\cdot\cdot$ &           14.5056 & \ion{Fe}{  xviii} &        $1s^22s^22p_{1/2}^22p_{3/2}^23d_{3/2}\; \rightarrow \;2s^22p^5~^2P_{3/2}$ &   6.80 &              \multicolumn{2}{c}{$\cdot\cdot\cdot$} \\
 10c & $\cdot\cdot\cdot$ &           14.5340 & \ion{Fe}{  xviii} &                   $2s^22p^4(^3P)3d~^2F_{5/2}\; \rightarrow \;2s^22p^5~^2P_{3/2}$ &   6.90 &              \multicolumn{2}{c}{$\cdot\cdot\cdot$} \\
 10d & $\cdot\cdot\cdot$ &           14.5710 & \ion{Fe}{  xviii} &                   $2s^22p^4(^3P)3d~^4P_{3/2}\; \rightarrow \;2s^22p^5~^2P_{3/2}$ &   6.90 &              \multicolumn{2}{c}{$\cdot\cdot\cdot$} \\
  11 &             15.01 &           15.0140 & \ion{Fe}{   xvii} &                       $2s^22p^5(^2P)3d~^1P_{1}\; \rightarrow \;2s^22p^6~^1S_{0}$ &   6.70 &                                        19.5 &  3.3 \\
 12a &             15.17 &           15.1760 & \ion{O }{   viii} &                                      $4p~^2P_{3/2}\; \rightarrow \;1s~^2S_{1/2}$ &   6.50 &                                         7.0 &  2.5 \\
 12b & $\cdot\cdot\cdot$ &           15.1765 & \ion{O }{   viii} &                                      $4p~^2P_{1/2}\; \rightarrow \;1s~^2S_{1/2}$ &   6.50 &              \multicolumn{2}{c}{$\cdot\cdot\cdot$} \\
 12c & $\cdot\cdot\cdot$ &           15.1980 & \ion{Fe}{    xix} &                    $1s^22s2p_{1/2}^22p_{3/2}^23s\; \rightarrow \;2s2p^5~^3P_{2}$ &   6.90 &              \multicolumn{2}{c}{$\cdot\cdot\cdot$} \\
  13 &             15.28 &           15.2610 & \ion{Fe}{   xvii} &                       $2s^22p^5(^2P)3d~^3D_{1}\; \rightarrow \;2s^22p^6~^1S_{0}$ &   6.70 &                                         3.7 &  2.3 \\
 14a &             15.96 &           16.0040 & \ion{Fe}{  xviii} &                   $2s^22p^4(^3P)3s~^2P_{3/2}\; \rightarrow \;2s^22p^5~^2P_{3/2}$ &   6.80 &                                        12.0 &  3.9 \\
 14b & $\cdot\cdot\cdot$ &           16.0055 & \ion{O }{   viii} &                                      $3p~^2P_{3/2}\; \rightarrow \;1s~^2S_{1/2}$ &   6.50 &              \multicolumn{2}{c}{$\cdot\cdot\cdot$} \\
 14c & $\cdot\cdot\cdot$ &           16.0067 & \ion{O }{   viii} &                                      $3p~^2P_{1/2}\; \rightarrow \;1s~^2S_{1/2}$ &   6.50 &              \multicolumn{2}{c}{$\cdot\cdot\cdot$} \\
 15a &             16.02 &           16.0710 & \ion{Fe}{  xviii} &                   $2s^22p^4(^3P)3s~^4P_{5/2}\; \rightarrow \;2s^22p^5~^2P_{3/2}$ &   6.80 &                                        19.8 &  4.3 \\
 15b & $\cdot\cdot\cdot$ &           16.1100 & \ion{Fe}{    xix} &              $1s^22s^22p_{1/2}2p_{3/2}^23p_{1/2}\; \rightarrow \;2s2p^5~^3P_{2}$ &   6.90 &              \multicolumn{2}{c}{$\cdot\cdot\cdot$} \\
  16 &             16.74 &           16.7800 & \ion{Fe}{   xvii} &                       $2s^22p^5(^2P)3s~^1P_{1}\; \rightarrow \;2s^22p^6~^1S_{0}$ &   6.70 &                                        13.3 &  2.8 \\
 17a &             17.06 &           17.0510 & \ion{Fe}{   xvii} &                       $2s^22p^5(^2P)3s~^3P_{1}\; \rightarrow \;2s^22p^6~^1S_{0}$ &   6.70 &                                        22.5 &  3.6 \\
 17b & $\cdot\cdot\cdot$ &           17.0960 & \ion{Fe}{   xvii} &                       $2s^22p^5(^2P)3s~^3P_{2}\; \rightarrow \;2s^22p^6~^1S_{0}$ &   6.70 &              \multicolumn{2}{c}{$\cdot\cdot\cdot$} \\
  18 &             17.58 &           17.6230 & \ion{Fe}{  xviii} &                          $2s^22p^43p~^2P_{3/2}\; \rightarrow \;2s2p^6~^2S_{1/2}$ &   6.80 &                                         6.7 &  2.4 \\
  19 &             18.66 &           18.6270 & \ion{O }{    vii} &                                      $1s3p~^1P_{1}\; \rightarrow \;1s^2~^1S_{0}$ &   6.30 &                                         5.7 &  2.8 \\
 20a &             18.96 &           18.9671 & \ion{O }{   viii} &                                      $2p~^2P_{3/2}\; \rightarrow \;1s~^2S_{1/2}$ &   6.50 &                                       119.7 &  6.8 \\
 20b & $\cdot\cdot\cdot$ &           18.9725 & \ion{O }{   viii} &                                      $2p~^2P_{1/2}\; \rightarrow \;1s~^2S_{1/2}$ &   6.50 &              \multicolumn{2}{c}{$\cdot\cdot\cdot$} \\
  21 &             21.60 &           21.6015 & \ion{O }{    vii} &                                      $1s2p~^1P_{1}\; \rightarrow \;1s^2~^1S_{0}$ &   6.30 &                                        39.8 &  6.9 \\
  22 &             21.80 &           21.8036 & \ion{O }{    vii} &                                      $1s2p~^3P_{1}\; \rightarrow \;1s^2~^1S_{0}$ &   6.30 &                                         6.4 &  3.9 \\
  23 &             22.08 &           22.0977 & \ion{O }{    vii} &                                      $1s2s~^3S_{1}\; \rightarrow \;1s^2~^1S_{0}$ &   6.30 &                                        24.5 &  5.5 \\
 24a &             24.79 &           24.7792 & \ion{N }{    vii} &                                      $2p~^2P_{3/2}\; \rightarrow \;1s~^2S_{1/2}$ &   6.30 &                                        20.8 &  4.3 \\
 24b & $\cdot\cdot\cdot$ &           24.7846 & \ion{N }{    vii} &                                      $2p~^2P_{1/2}\; \rightarrow \;1s~^2S_{1/2}$ &   6.30 &              \multicolumn{2}{c}{$\cdot\cdot\cdot$} \\
 25a &             33.73 &           33.7342 & \ion{C }{     vi} &                                      $2p~^2P_{3/2}\; \rightarrow \;1s~^2S_{1/2}$ &   6.10 &                                        22.5 &  4.8 \\
 25b & $\cdot\cdot\cdot$ &           33.7396 & \ion{C }{     vi} &                                      $2p~^2P_{1/2}\; \rightarrow \;1s~^2S_{1/2}$ &   6.10 &              \multicolumn{2}{c}{$\cdot\cdot\cdot$} \\
\end{longtable}
\normalsize

\end{document}